\documentclass{naturesjs}
\pdfoutput=1
\usepackage{hyperref}
\usepackage{graphicx}
\usepackage{epsfig}
\usepackage[font={small}]{caption}
\usepackage{tablefootnote}
\usepackage{amssymb}
\usepackage{amsmath}
\usepackage{color}
\usepackage{comment}
\usepackage{xparse,soul}
\usepackage{gensymb}

\title{A three-dimensional map of the hot Local Bubble using diffuse interstellar bands}

\author{
Amin Farhang$^{1,2}$,
Jacco Th. van Loon$^{3}$,
Habib G. Khosroshahi$^{1}$,
Atefeh Javadi$^{1}$
\&
Mandy Bailey$^{4}$
}

\begin{document}
\maketitle

\begin{affiliations}
 \item School of Astronomy, Institute for Research in Fundamental Sciences, 19395--5531 Tehran, Iran
 \item Department of Physics and Astronomy, The University of Western Ontario, N6A 3K7, Canada
 \item Lennard-Jones Laboratories, Keele University, ST5 5BG, UK
 \item The Open University, Associate Lecturer Services (STEM), 351, Altrincham Road, Sharston, Manchester, M22 4UN, UK

\end{affiliations}

\begin{abstract}
The Solar System is located within a low-density cavity, known as the Local Bubble\cite{1987ARA&A..25..303C,2010A&A...510A..54W,2014A&A...561A..91L}, which appears to be filled with an X-ray emitting gas at a temperature of 10$^6$ K\cite{2014Natur.512..171G}. Such conditions are too harsh for typical interstellar atoms and molecules to survive\cite{2010A&A...510A..54W,2014A&A...561A..91L}. There exists an enigmatic tracer of interstellar gas, known as Diffuse Interstellar Bands (DIB), which often appears as absorption features in stellar spectra\cite{2000ApJ...537..690H, 2013ApJ...779...38P,2015A&A...576L...3M}. The carriers of these bands remain largely unidentified\cite{2006JMoSp.238....1S}. Here we report the three-dimensional structure of the Local Bubble using two different DIB tracers ($\lambda$5780 and $\lambda$5797), which reveals that DIB carriers are present within the Bubble\cite{2016A&A...585A..12B,2015ApJ...800...64F,2015ApJS..216...33F}. The map shows low ratios of $\lambda$5797/$\lambda$5780 inside the Bubble compared to the outside. This finding proves that the carrier of the $\lambda$5780 DIB can withstand X-ray photo-dissociation and sputtering by fast ions, where the carrier of the $\lambda$5797 DIB succumbs. This would mean that DIB carriers can be more stable than hitherto thought, and that the carrier of the $\lambda$5780 DIB must be larger than that of the $\lambda$5797 DIB\cite{2010A&A...510A..37M}. Alternatively, small-scale denser (and cooler) structures that shield some of the DIB carriers must be prevalent within the Bubble, implying that such structures may be an intrinsic feature of supernova-driven bubbles.
\end{abstract}

The origin of the Local Bubble (LB) is unknown, but measurements of $^{60}$Fe column densities can be explained by successive explosions of massive stars (supernovae) within the Scorpius--Centaurus stellar group\cite{2006MNRAS.373..993F}. The high temperature within the cavity is hostile to atoms, molecules and dust grains\cite{2010A&A...510A..54W,2014A&A...561A..91L,2014Natur.512..171G}. Observations aimed at detecting highly ionized gas, which could be present at million-degree temperatures, reveal no such plasma within the LB\cite{1999ApJ...517..850H}. The DIB carriers offer an alternative tracer of the wall and interior of the LB. Although their nature is unknown, recent studies indicate that DIB carriers are likely carbon-based organic molecules\cite{2006JMoSp.238....1S}. They are universal and have been detected in different environments of Milky Way (MW) and within different galaxies\cite{2000ApJ...537..690H,2013ApJ...779...38P,2015A&A...576L...3M} and in earlier partial results from our survey\cite{2016A&A...585A..12B,2015ApJ...800...64F,2015ApJS..216...33F}. Our 3D map of the distribution of DIB carriers within the vicinity of the Sun opens a new window on a substantial fraction of interstellar carbon that may be locked up in the molecular carriers and that may play an important role in interstellar chemistry; this may include the C$_{60}^{+}$ (``buckyballs'') anion which may be responsible for the DIBs at $9577$ \AA\ and $9632$ \AA\ wavelengths\cite{Fulara93}. Therefore, we employed two of the strongest DIB tracers at $5780$ \AA\ and $5797$ \AA\ wavelengths, in direction of 359 different sightlines, to mapped the LB out to a distance of $200$ pc. While the $\lambda5780$ DIB could possibly persist under harsh conditions of the LB\cite{1991PASP..103.1005S, 2011A&A...533A.129V}, the $\lambda5797$ DIB cannot survive in such an environment and must be shielded within the inner regions of clouds\cite{2015ApJ...800...64F}.

The outlines of the LB have been mapped with absorption lines of singly ionized calcium (Ca\,{\sc ii}) and neutral sodium (Na\,{\sc i}) in the spectra of background stars\cite{2010A&A...510A..54W}, as well as by the attenuation of stellar light by interstellar dust based on $E(B-V)$ measurements\cite{2014A&A...561A..91L,2017A&A...606A..65C,2019arXiv190204116L}, revealing that the LB cavity extends out to 80 pc in the Galactic plane (GP) and up to several hundred pc perpendiculars to the GP. Some dust maps have been produced by translating the strength of $\lambda$15273 DIB into $E(B-V)$ and merging it with other $E(B-V)$ measurements as priors; thus not strictly speaking a DIB map, but really a dust map \cite{2017A&A...606A..65C}. Also, a pseudo-3D map of the $\lambda$8620 DIB within 3 kpc from the Sun have been produced recently\cite{2014Sci...345..791K}. They compared the DIB map with the dust distribution and found the $\lambda$8620 DIB and dust to have a similar distribution, but the scale height of the DIB exceeds that of the dust. This is a clear indication that DIBs and dust do not necessarily trace the same interstellar material and/or conditions\cite{2016A&A...585A..12B}. Their pseudo-3D map traces large-scale structure, with the whole LB being confined to one voxel.

Because of the high temperature of the LB, it is difficult to detect ordinary atoms within the LB. However, our observations show that DIB carriers are present within the LB\cite{2016A&A...585A..12B,2015ApJ...800...64F,2015ApJS..216...33F}. An example of different DIBs within and around the LB is shown in Fig.\,1. To ensure that the observed DIBs are physical features not noise, only those absorptions with a confidence level exceeding $3\sigma$ are accepted in the mapping. In the following, we present three principal slices of the 3D distribution of the DIB pseudo volume density within $200$ pc from the Sun. The maps are in Galactic coordinates, i.e. the Sun is located at the origin and the primary direction is towards the Galactic Center (GC). The slices show the maps in imaginary planes of the Galactic, meridian and rotational planes. The GP slices the MW disk, the meridian plane is perpendicular to the GP with its x-axis pointing towards the GC, and the rotational plane is perpendicular to the GP and faced towards the GC.

In upper left panel of Fig.\,2 ($\lambda5780$ in the GP) the bulk of the material lies in front of the Scutum and Aquila dark nebulae (in the 1st quadrant in the upper-right sector) at distances of 30--120 pc from the Sun. The outer layer of this dense DIB structure coincides with a dense concentration in the dust map, however, the inner region is stretched within the LB and continued to interrupt the whole LB where there is not any dust\cite{2014A&A...561A..91L} (see Supplementary Fig.\,1 for an RGB map of dust, DIB and Na\,{\sc i}). A lower-density DIB filament lies in front of the Cygnus rift molecular clouds. In this direction, X-rays with energies of order keV probably affect the DIB carriers beyond a distance of approximately $100$ pc\cite{2005A&A...438..187C}. In the 2nd quadrant (upper-left sector) DIB carriers are abundant in the direction of the Taurus dark clouds and molecular. The 3rd quadrant (lower-left sector) is characterized by a general paucity of the DIB carrier, in the direction of the $\beta$ Canis Majoris (CMa) interstellar tunnel and the GSH 238+00+09 supershell ($l=260$\degree). However, a low-density DIB trunk defies the odds; it may be associated with the photo-ionizing effect of $\beta$CMa \cite{2010A&A...510A..54W}. Finally, the 4th quadrant (lower-right sector) features the most famous cavity connected to the LB (Loop I) in the direction of the GC ($l=345$\degree) at a distance of $200$ pc. A narrow tunnel connects the LB with Loop I, as revealed in previous LB maps\cite{2010A&A...510A..54W,2014A&A...561A..91L}; our map reveals that it is filled with a DIB filament and that Loop I itself is filled with DIB material as well (see Supplementary Fig.\,2 for more slices).

The vertically extended structures are seen in the meridian plane (middle left panel of Fig.\,2), one towards the GC in front of the Ophiuchus and Lupus complexes and R Coronae Australis (CrA) and another in the opposite direction in front of the Taurus star formation complex in between the Hyades and Pleiades star clusters. The atomic maps of the LB in this view show a striking open-ended tunnel known as the Local Chimney, tilted at an angle of 35$\degree$ from the North Galactic Pole (NGP)\cite{1999A&A...352..308W,2010A&A...510A..54W}, but our map shows that the tunnel is not entirely devoid of DIB carriers. Interestingly, there are some high latitude clouds (H\,{\sc i} shells) in this direction \cite{2012A&A...545A..21P}. The tilted tunnel can also be discerned in the rotational plane, but most noticeable is the DIB complex in the 3rd quadrant and material some $150$ pc below the GP in the direction of the South Galactic Pole.

In the right column of Fig.\,2, we present the $\lambda5797$ distributions. The carrier of the $\lambda5797$ DIB is highly susceptible to the destructive effect of energetic photons\cite{2016A&A...585A..12B,2015ApJ...800...64F}. The GP view shows the largest concentrations lie between $30$--$120$ pc distance in the direction of Scutum and a filament extends over $200$ pc to meet Loop I, both akin to what was seen in the $\lambda5780$. Likewise, the Local Chimney is visible in the meridian plane and a thin filament coincides with the densest $\lambda5780$ DIB concentration in the 3rd quadrant of the rotational plane. Apart from the similarity between the two DIB carriers there are also notable differences, with the $\lambda5797$ DIB distribution generally more fragmented and more tenuous. The $\lambda5780$ DIB is known to trace relatively energetic environments as compared to the $\lambda5797$ that traces more neutral, shielded regions (see Supplementary Fig.\,3 where the strength of the $\lambda5780$ saturates at high Na\,{\sc i} densities). Indeed, the ratio of the two DIB carrier distributions shows that the interior of the LB is depleted in the $\lambda5797$ DIB carrier, more so than the $\lambda5780$ DIB carrier (Fig.\,3) though more neutral, shielded cloudlets seem to persist within the LB.

The occurrence of DIBs within the LB, Local Chimney and the pathway tunnels, emphasizes that the $\lambda5780$ and $\lambda5797$ DIBs cannot all trace dust perfectly, as some studies showed that different DIBs trace different parts of clouds\cite{2016A&A...585A..12B}. Conversely, it implies that at least some DIBs will yield different information compared to dust maps.

The question arises how DIBs could survive in such a high temperature? To address this question, we should emphasize that, in general, two different processes could destroy atoms and molecules in the hot regions of the ISM. Photo-dissociation by X-ray photons from stellar winds and supernovae, and sputtering by ions and electrons in the hot plasma and behind fast non-radiative shocks. Both processes can lead to the loss of atoms from large, carbonaceous molecules (like PAH and DIBs) and causing molecular destruction. On the other hand, to better understand the LB environment, we could compare the LB with the Orion nebula and the M82 galaxy. The center of Orion nebula (Orion-S), has a moderate temperature (T = $10^{4}$ K) measured from [N\,{\sc ii}], and a high electron density (n$_{e} = 10^{4}$ cm$^{-3}$) estimated from the [S\,{\sc ii}] forbidden doublet-line ratio\cite{2015A&A...582A.114W}, while the starburst M82 galaxy shows a high temperature (T = 5.8 $\times$ 10$^{^6}$ K) and low density (n$_{H}$ = 0.013 cm$^{−3}$) in its bipolar outflow\cite{2008MNRAS.386.1464R}. Therefore, the LB (with n$_{H}$ = 0.01 cm$^{−3}$ and T=10$^{6}$ K) is more similar to M82 but the small colder regions within the LB (such as three sub-regions of Ca\,{\sc ii}) could be more akin the Orion Nebula. In the galactic outflows from M82, the electron collisions destroyed PAHs with 50--200 C-atom, although the bigger molecules can survive longer, eventually, all PAHs are destroyed within several thousand years\cite{2010A&A...510A..37M}. In the denser and colder regions of the Orion Nebula, the PAH erosion is not caused by electron sputtering. In these regions the ion (He) collisions damage the PAHs, yielding a timescale of 10$^7$ years for the PAH destruction\cite{2010A&A...510A..37M}.

Based on this discussion, in a high-temperature and low-density gas, small  PAHs would be destroyed. Therefore, the survival of small molecules in such conditions requires a protective environment and/or an efficient reformation mechanism. For the small shielded regions of the LB's interior, the DIBs can survive in the surface layers of cold clouds within the hot LB for several million years (similar to Orion)\cite{2010A&A...510A..37M}. However, for non-shielded regions of high temperature, a possible reason for observing DIBs could be described with a formation mechanism. As some very low reddening is still seen within the LB (with E(B-V) $<$ 0.1 mag), a small fraction of interstellar grains \cite{2012ApJ...761...35M} has still remained within the LB. When these grains are eroded by electron collisions, any DIB carriers stuck on them are released into the ISM. Therefore, we could observe weak DIBs within the LB after $\sim$14 Myr (estimated age of LB)\cite{2006MNRAS.373..993F} showing that before the creation of the LB, this part of the ISM was characterized by dense, cool gas. Therefore during the lifetime of the LB, DIB carriers have been released into the ISM at a steady rate, making it possible to see very weak DIBs within the LB to this day.

The evolution of multiple supernova shells is expected to instantiate dynamical instabilities within the LB. These have been linked to the formation of local cool clouds\cite{Breitschwerdt06} while tunnels and holes in the wall may have resulted from the Vishniac instability\cite{2013MNRAS.435.3600P}. Cold, dense structures may also form out of hot gas through thermal instabilities\cite{2015MNRAS.449.1057G}. It thus appears plausible from our maps that DIBs such as the $\lambda5780$ may be sensitive, and potentially unique tracers of such structures at the boundary between cool and hot ISM phases.

Our maps show the LB in a truly new light -- the DIBs likely highlight the cloud surfaces and shallow clouds that are in direct contact with energetic electrons and/or photons, while the densest interstellar structures are better mapped with atomic species or dust (see Supplementary Fig.\,1). Thus, the same supernovae that only a few million years ago may have given rise to the global increase in the $^{60}$Fe isotope found in deep-sea crusts, and which occurred at the same time as when Earth's temperature started to decrease\cite{2016Natur.532...69W}, may have descended upon Earth, or may do so in the future as Earth traverses through the walls of the LB.

%

\clearpage
\begin{addendum}

\item[Correspondence]
Correspondence and requests for materials should be addressed to A. Farhang~(email: a.farhang@ipm.ir)

\item[Acknowlegdements]
We wish to thank the Iranian National Observatory (INO) and School of Astonomy at IPM for facilitating and supporting this project for the Norhern part of the observations and Keele University for their hositality during A.F. and A.J. visits, and for their support of the Souhern observations. Also, we wish to thank all of the staff at La Silla in hile and the ING staff at La Palma, Spain -- scientific, technical and admn -- for their support. M.B. acknowledges an STFC studentship at Keele Uniersity. This research has made use of the SIMBAD database, operated at CDS Strasbourg, France. A.F would like to thanks Dr. Jan Cami for his coments on the paper. All authors deeply thanks the editor for his constructive comments and the anonymous referees for their valuable comments.

\item[Author Contributions]
The main idea for this work was proposed by J.v.L. and M.B. The Northern observations (with the Isaac Newton Telescope - INT) were proposed by A.J. and carried out by A.F. and H.K. as well as other colleagues within the School of Astronomy at IPM, while M.B. and J.v.L. proposed and observed the Southern targets (with the New Technology Telescope - NTT). A.F. performed the data analysis and data reduction of the Northern observations, while M.B. did the same for the Southern sample. A.F. implemented the inverse method on the combined data set and together with J.v.L. and H.K. wrote the manuscript. All authors read and commented on the manuscript and contributed to the scientific interpretation.

\item[Author Information]
A. Farhang - Present address: Institude in research in fundamental science, Tehran, Iran. J. Th. van Loon - Present address: Lennard-Jones Laboratories, Keele University, ST5 5BG, UK. H. Gh. Khosroshahi - Present address: Institude in research in fundamental science, Tehran, Iran. A. Javadi -
Present address: Institude in research in fundamental science, Tehran, Iran. M. Bailey - Present address: The Open University, Associate Lecturer Services (STEM), 351, Altrincham Road, Sharston, Manchester, M22 4UN, UK.

\end{addendum}

\clearpage

\begin{figure*}
\centering
\includegraphics[width=14cm]{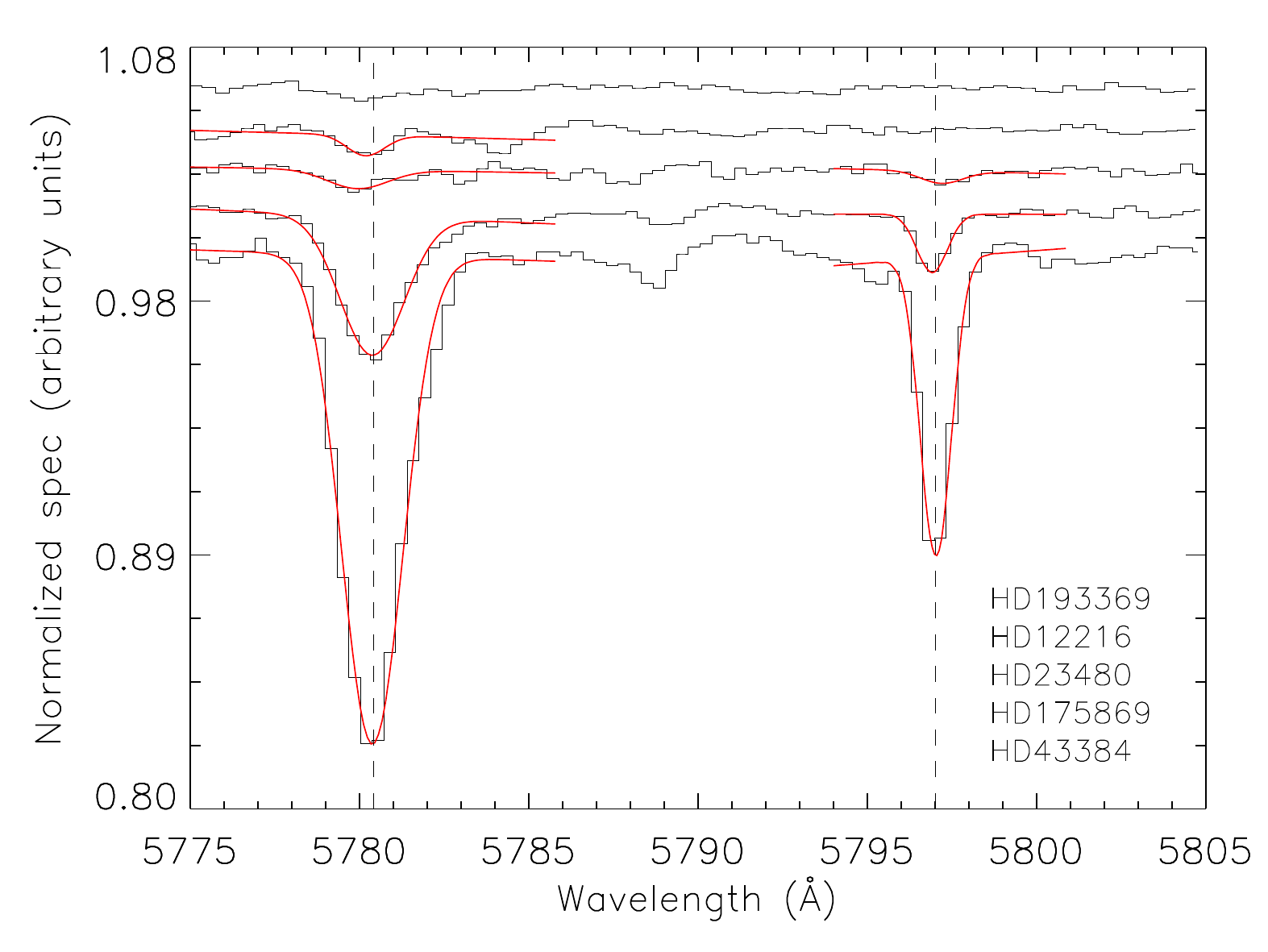}
\label{fig:f1}
\caption{\textbf{Observed DIB specta within and around the LB}. Normalized high S/N spectrum of strong (HD 43384 with EW$_{\lambda5780} = 452 \pm 48$ m\AA, EW$_{\lambda5797} = 125 \pm 15$ m\AA), mid (HD 175869 with EW$_{\lambda5780} = 135 \pm 16$ m\AA, EW$_{\lambda5797} = 29 \pm 5$ m\AA), and weak (HD 23480 with EW$_{\lambda5780} = 16 \pm 4$ m\AA, EW$_{\lambda5797} = 8 \pm 3$ m\AA) DIBs as well as a sightline with absence of $\lambda5797$ (HD 12216 with EW$_{\lambda5780} = 17 \pm 3$ m\AA) and an example of sightlines with no DIBs at all (HD193369). The red lines show Gaussian fits to the DIBs. All spectra are moved to interstellar rest frame. The HD 12216 (at $d=44$ pc) is an example of DIBs located within the LB, and HD 23480 is located at the wall of the LB (at $d=105$ pc), while HD 43384 is located well outside the LB at $d=641$ pc.}
\end{figure*}
\clearpage

\begin{figure*}
\centering
\includegraphics[width=14cm]{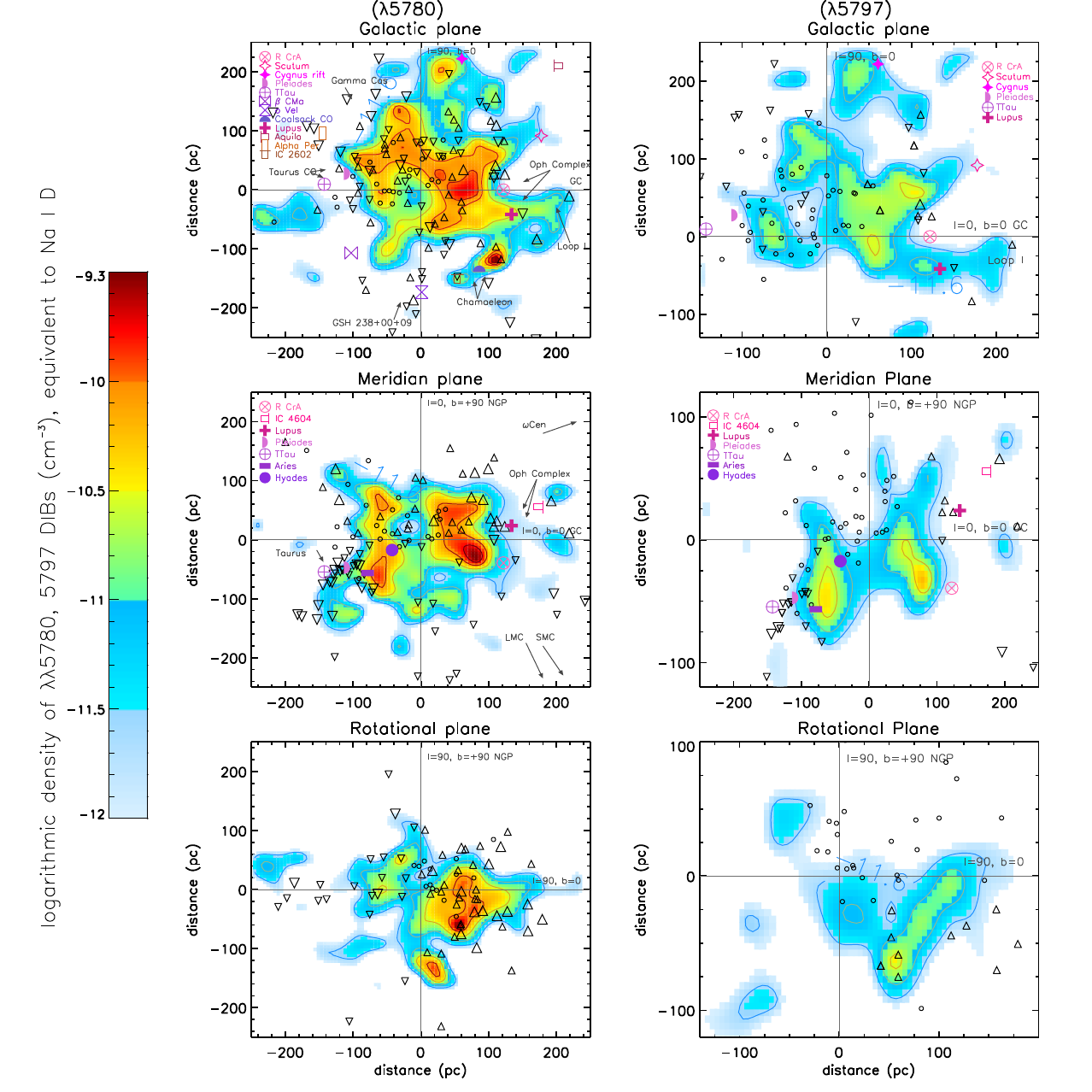}
\label{fig:f2}
\caption{\textbf{$\lambda5780$ DIB distribution in three principal slices}. The quantities are colored based on the logarithmic volume densities of the equivalent amount of neutral sodium, with redder regions tracing denser parts and bluer regions the rarefied mediums. Blue, gold, red and dark red contours correspond to $\log n$ (cm$^{-3}$) = $-11.6$, $-10.6$, $-10.1$ and $-9.7$. The positions of nearby nebulae and star clusters are plotted with various symbols. Triangles represent the projection of observed stars with distances less than $30$ pc to this particular slice: upward if located above the GP and downward if below it, and with the size proportional to the derived column density. Open circles are sightlines with zero DIB column densities assigned if the standard deviation within $\pm3$ \AA\ around the DIB position equaled the noise. The Sun is located in the center of the map, the name of each slice is printed in panel title. The distance scale is in units of parsec.}
\end{figure*}
\clearpage

\begin{figure*}
\centering
\includegraphics[width=14cm]{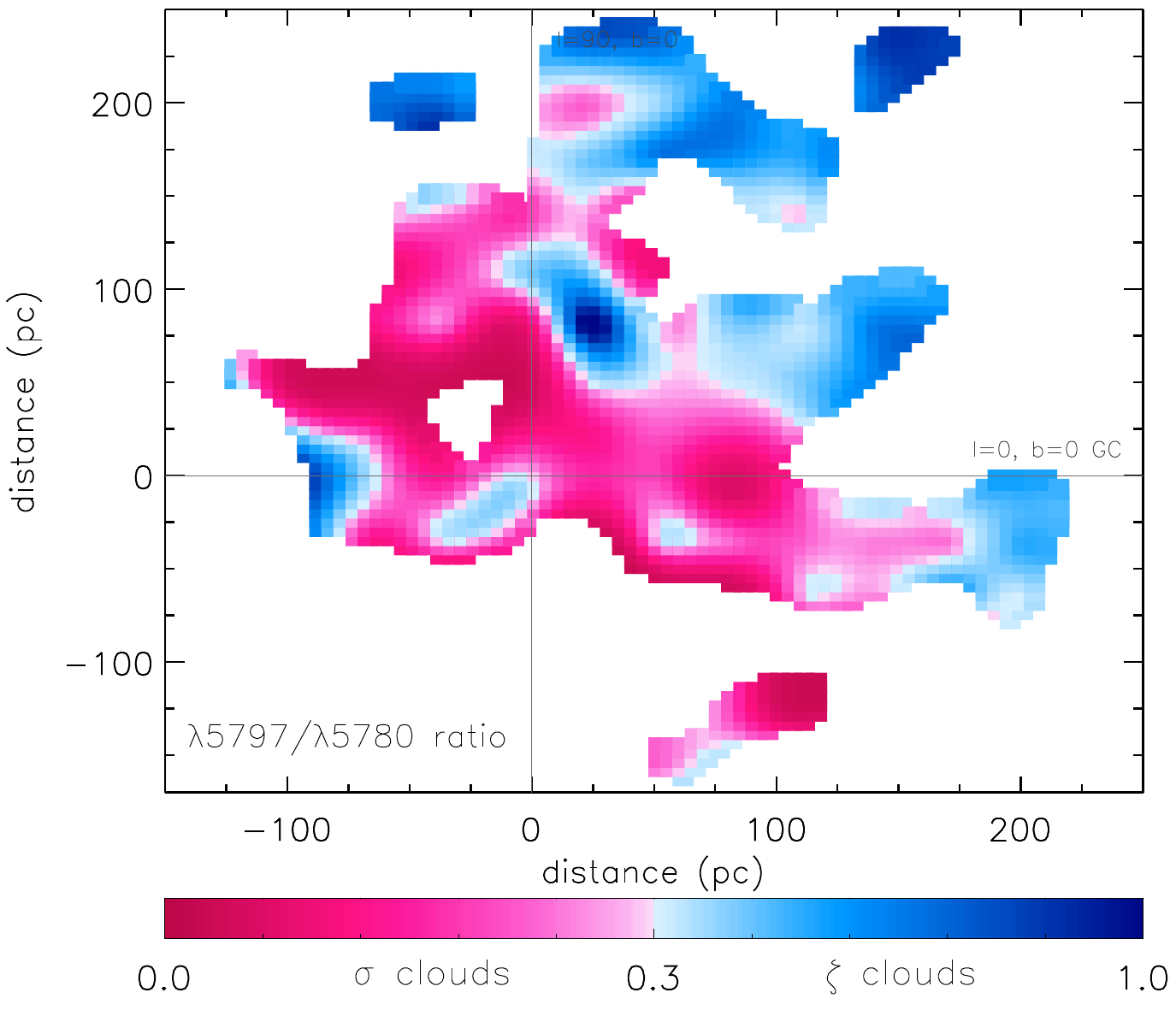}
\label{fig:f3}
\caption{\textbf{$\zeta$ and $\sigma$ cloud distribution within the GP}. The ratio of $\lambda5797/\lambda5780$ DIB equivalent width is thought to probe the UV radiation field. The $\sigma$ sightlines, where have $W(5797)/W(5780) < 0.3$, sample regions with high UV intensity as the $\lambda5797$ DIB carrier is suppressed while the $\lambda5780$ DIB carrier is enhanced, and typically probe the envelopes of clouds. The $\zeta$ sightlines, where $W(5797)/W(5780) > 0.3$, sample regions where the $\lambda5797$ DIB carrier is protected from high energy photons and the $\lambda5780$ DIB carrier is suppressed, and typically probe the interiors of clouds (but not the highest densities). As is clear from the map, the LB and the passage towards Loop I are filled with $\sigma$ clouds due to the harsh environment, but still some small $\zeta$ clouds can be found immersed within it.}
\end{figure*}
\clearpage

\clearpage
\begin{methods}

\subsection{Observations}
To map DIB absorption in and around the LB, we conducted a high signal-to-noise (S/N) survey of 637 nearby early-type stars in both hemispheres (see Supplementary Fig.\,4). The Southern hemisphere survey was carried out with the 3.5m New Technology Telescope (NTT) at La Silla, Chile\cite{2016A&A...585A..12B}, and the Northern survey has been conducted with 2.5m Isaac Newton Telescope (INT) at La Palma, Spain\cite{2015ApJ...800...64F}. The Southern observations were performed during 10 nights from March 2011 to August 2012. The Faint Object Spectrograph and Camera (EFOSC2) was used at a spectral resolving power of $R=\lambda/\Delta\lambda = 5500$, and covering a wavelength range from 5672 \AA\ to 6772 \AA. The Northern observations were carried out over the 35 nights from October 2011 June 2013. In these observations we used the Intermediate Dispersion Spectrograph (IDS) to cover the $\lambda\lambda5780$, $5797$ and $5850$ DIBs in the $5750$--$6040$ \AA\ region at a spectral resolving power of $R=\lambda/\Delta\lambda = 2000$.

More than 60 flat frames (captured by quartz lamp), about 15 arc lamp exposures (CuAr+CuNe at the INT and He+Ar at the NTT), and a large number of bias frames were taken for each observing night. To obtain a typical $S/N>1000$, we observed 9--25 science frames for each target. For data reduction, first, we produced master frames of bias and flat field by combining their individual spectra. The master bias then has been subtracted from master flat and science frames, and finally, the science frames were divided by the master flat. Later the weighted optimal extracted spectra were calibrated using arc frames. Possible bright cosmic ray impacts on detector have been removed using the sigma-clipping method (see Supplementary Fig.\,5 for a comparison between measurements for a select number of targets that had been observed both at the INT and NTT).

\subsection{Measurements}
Quantifying the equivalent widths (EW) for DIB features is extremely challenging, especially when they are weak, shallow, and/or blended with other spectral lines. Since the DIB profile is probably composed of an overlapping set of an unknown number of transitions, the overall shape of their profiles is unexplained (and differs between DIBs). Single cloud line-of-sights confirm that most of them do not have a perfect Voigt profile. For instance, high-resolution spectra of the $\lambda$5797 profile reveal substructure. Because we are using low-resolution spectrographs, the shape of $\lambda$5780 and $\lambda$5797 are matched by a Gaussian profile\cite{2009MNRAS.399..195V}. First, a normalized, rectified spectrum is constructed by fitting a low order Legendre polynomial, by choosing $\pm$1--2 \AA\ around the central DIB and masking the DIB feature itself (note that in figure 1 we only plot a small range of the continuum in order to highlight the fit to the DIB profile). Then, the line widths of $\lambda$5780 and $\lambda$5797 DIBs were obtained in terms of the standard deviation ($\sigma$) of Gaussian fits,  and the FWHM calculated as FWHM = 2(2ln 2)$^{1/2}\sigma$ = 2.355$\sigma$. The EW is then calculated from the integral of the Gaussian fit.

The major source of uncertainty in DIB's EW is their blending with other stellar and interstellar lines. For instance, the $\lambda$5780 DIB could be blend with $\lambda$5778 DIB feature\cite{1975ApJ...196..129H}. For each DIB, the statistical-uncertainty computed by summing the standard deviation of the Gaussian fit's residuals\cite{2009MNRAS.399..195V, 2011A&A...533A.129V}. However, the main source of EW uncertainty in DIB studies is the systematic-error of continuum position. This error was estimated based on fitting three different continuum lines in $\pm$2 \AA\ range around the peak (linear fit, quadratic fit, and simultaneous fit to the DIB and a linear continuum\cite{2013ApJ...774...72K}). Later, the interval between the highest and lowest values of EW was taken as the systematic-uncertainty.

\subsection{Data selection}
We used high-quality data from the South and North DIB surveys with absorptions stronger than $3 \sigma$ (where $\sigma$ is the standard deviation of the noise). Cool stars, with spectral types later than A3, were omitted to avoid contamination from stellar spectral lines. The latter was further mitigated by only using those sightlines with equivalent spectral width exceeding $6$ m\AA\ and DIB full width at half maximum (FWHM) exceeding $0.5$ \AA. On the other hand, we consider sightlines with no discernible feature at the position of the $\lambda5780$ and $\lambda5797$ DIBs as zero column density sightlines. The absence of DIB features in these sightlines shows that the DIB density is negligible in some areas and moreover helps to reveal the fragmental distribution of DIBs in the LB. Also, since the spectrum of binary systems is contaminated by the stellar spectra, to avoid any non-real estimation, we reject those in our final catalog. Based on these selection criteria, among 637 observed targets only 359 sightlines used for mapping the LB. The measured DIB equivalent widths can be described by a density which decreases exponentially with distance from the GP, with a scale height of $100$ pc\cite{2010A&A...510A..54W,2014A&A...561A..91L}. This can be compared to the scale height of Na\,{\sc i}, $h_{0} = 170$ pc, and that of Ca\,{\sc ii}, $h_{0} = 450$ pc\cite{2010A&A...510A..54W}. The equivalent width can be converted to column density by the following equation:
\begin{equation}
    N=1.13 \times 10^{17} \times \frac{W}{f \lambda ^{2}}
\end{equation}
Here, the wavelength is in \AA, $W$ is the equivalent width in m\AA, and $f$ is the oscillator strength. The structure of the carrier must be known to be able to compute the oscillator strength but the DIB carriers are unknown. The $\lambda5797$ DIB carriers tend to be present in the dense core of neutral Na\,{\sc i} clouds, protected from the UV background radiation\cite{2015ApJ...800...64F}, while the $\lambda5780$ DIB carriers are more abundant in the skin of the Na\,{\sc i} clouds. On the other hand, both the $\lambda5780$ and $\lambda5797$ DIBs exhibit a strong correlation with Na\,{\sc i}, understood as a general gas column dependency\cite{2016A&A...585A..12B,2015ApJ...800...64F}. Therefore, in order to be able to assign a value to the oscillator strength of the DIBs, we suppose that the DIB column densities are in the same range as the Na\,{\sc i} column densities. By taking $f = 0.3$ the Na\,{\sc i} and $\lambda5780$ DIB column densities lie in the same range (as shown in Supplementary Fig.\,3). Thus we derived pseudo column densities for the carriers of the $\lambda5780$ and $\lambda5797$ DIBs.

All distances to the target stars are measured from parallaxes from the second Gaia data release (GDR2)\cite{2018A&A...616A..10G,2018A&A...616A...1G}. However, reliable distances for the majority of stars in the GDR2 cannot be obtained only by inverting the parallax. A correct inference procedure must instead be used to account for the nonlinearity of the transformation and the asymmetry of the resulting probability distribution considering the low signal-to-noise ratios of many measured parallaxes in the GDR2. Therefore we used the catalog of Bailer-Jones et al. (2018)\cite{2018AJ....156...58B} who provide purely geometric distance estimates for GDR2 sources.

\subsection{Inverse method}
\label{sec:invmethod}
In general, data values can be calculated directly from a given model (the `forward’ problem). On the other hand, to reconstruct a model from a set of measurements one must use an inverse method to estimate values at the positions where no direct measurements exist. It is possible to find the probability density of posterior estimates by assuming a Gaussian distribution\cite{1982RvGSP..20..219T} for both the data ($N$) and a priori model parameters ($m$). By defining the model parameters and data as a set of $X = [N,m]^{T}$ ($T$ is the transpose matrix) vectors, the a priori distribution and the model form a cloud in the parameter space. This cloud is centered on the observed data and the mean of a priori parameters. The shape of the cloud is described by the covariance matrix, $C_{X}$, which on its diagonal contains the covariance matrix of the data, $C_{obs}$, and the covariance of a priori values, $C_{prior}$\cite{1984GeoJ...76..299T}. The probability distribution takes the following form:
\begin{equation}
    \rho (X) = const.exp \left\{\frac{-1}{2} \left( X-X_{0}^{T} \right) C_{0}^{-1} \left( X-X_{0} \right)\right\}
    \label{eq:e1}
\end{equation}
One approach to estimating the optimal density distribution is to find the maximum likelihood point of $\rho(X)$ on the surface of $f(X)=0$. To estimate this likelihood point one should maximize $\rho(X)$ with the $f(X)=0$ constraint, while minimizing the exponential argument of $\rho(X)$, instead of determining the whole probability distribution\cite{1982RvGSP..20..219T}. This equation could be solved with Lagrange multiplier equations:
\begin{equation}
    \left[ X-X_{0} \right] =C_{X}G^{T} \left\{ GC_{X}G^{T} \right\} ^{-1} \left\{ G \left[ X-X_{0} \right] -f \left( X \right)  \right\}
\end{equation}
Where $G$ is the matrix of partial derivatives $G=\partial g \left( m \right)/{ \partial m}$. In the above equation, the variable $X$ appears on both sides and $f$ is a function of $X$; this makes it difficult to solve explicitly\cite{1982RvGSP..20..219T}. But the equation could be generalized in an iterative process, starting with some initial trial solution, $m_{0}$. If this a priori guess was close enough to the maximum likelihood point, then the successive approximations will converge to the true solution, $m_{est}$, within a few iterations.

In the case of constructing a 3D volume density distribution from a set of 2D column densities, $N = [N_{1}, N_{2}, ..., N_{n}]^{T}$, with an inverse method along different sightlines, the most probable likelihood point could be found with an iterative Newtonian method\cite{1982RvGSP..20..219T, 1984GeoJ...76..299T}. The volume density, $\rho$, is related to column density, $N$, by a simple integral equation, and the gas volume density in the ISM is known to approximately vary as a function of distance and Galactic latitude\cite{2010A&A...510A..54W, 2014A&A...561A..91L} as $\rho_{0}(r)\exp(-|rsin(b)|/h_{0})$. Hence, to parameterize $\rho$ in each point of space, $N$ can be expressed as follows:
\begin{equation}
    N=g \left( m \right) = \int _{0}^{r} \rho _{0}exp \left( m \left( r \right) -\frac{ \vert rsin \left( b \right)  \vert }{h_{0}} \right) dr
\end{equation}
Therefore, the observed column densities are related to volume densities by a model $N = g(m)$. To define the $g(m)$ function, the optimal $m$ parameter must be found with an inverse method. Using a Newtonian method one could estimate the optimal $m$ parameter with the following equations:
\begin{equation}
    \widetilde{m}_{k+1}=\widetilde{m}_{k}+C_{post} \left\{ G_{k}^{T} \left( C_{obs}^{-1} \left( N_{obs}-g \left( \widetilde{m}_{k} \right)  \right)  \right) ^{T}-C_{prior}^{-1} \left( \widetilde{m}_{k}-m_{0} \right)  \right\}
\end{equation}
\begin{equation}
    C_{post}= \left( G_{k}^{T}C_{obs}^{-1}G_{k}+C_{prior}^{-1} \right) ^{-1}
\end{equation}
Here, $m_{0}$ is the a priori value of the parameter, $G^{T}$ is the transpose operator of $G$, the subscript $k$ refers to the iteration order, and $\widetilde{m}_{k}$ is the estimate of the m parameter in the $k$-th iteration. The value $C_{prior}$ represents the covariance between point $r$ and $r'$, which in its simplest analytical form is thus\cite{1982RvGSP..20..219T}:
\begin{equation}
    C_{prior}= \sigma _{m}^{2}exp⁡ \left( \frac{-r^{2}+r^{'2}-2rr^{'}cos \left(  \theta  \right) }{ \zeta ^{2}} \right)
\end{equation}
Here, $\theta$ is the angle between the i-th and j-th sightline, $\zeta$ is the correlation length, and $\sigma$ is the a priori uncertainty for each point. At each iteration, a $\chi^{2}$ minimization criterion is used to control the algorithm convergence:
\begin{equation}
    \chi ^{2}= \sum _{i=1}^{n}\frac{ \left( N_{i}^{obs}-N_{i}^{model} \right) ^{2}}{ \sigma _{N_{i}}^{2}}
\end{equation}
The error on the observed column density ($\sigma_{N}$) is computed using similar methods as in previous works\cite{2010A&A...510A..54W,2014A&A...561A..91L}.

The smoothing length implies that spatial detail below this characteristic scale will be smoothed out. The smoothing length used in this study is $\zeta = 30$ pc which means that, if at a given point there is a deviation from the a priori model of given sign and magnitude, we want the deviation in the a priori model to be smoothed based on the column densities in the neighborhood of 30 pc boxes. The value for $\zeta$ was determined on the basis of the average distance between targets.

\subsection{Priors}
The inverse method is an ill-posed problem with a non-unique solution for each iteration as the estimated solution will depend on the initial guess. To determine the dependency of the distribution to the a priori guesses, we checked three different prior estimates. Firstly, we used a clumpy a priori guess for the initial value. We chose a clump as dense as the observed column density and change its position along the sightline at a fraction of $0.25$, $0.5$, and $0.75$ of the star’s distance. By using these prior values the iteration could not converge to an acceptable $\chi^{2}$ (the best-achieved $\chi^{2}=12.4$). In the second attempt, we divided each sightline into several equal segments and for each of those adopted the same initial value of the $m_{0}$ parameter based on the measured column density. With this initial guess, the inversion converged to $\chi^{2}=2.67$. However, some small dense clouds accumulated near the Sun despite a general accumulation of zero column density sightlines at these short distances. In the final attempt, we used an a priori Gaussian distribution centered at the middle of the sightline with the density distributed around $m_{0}$ (estimated based on the measured column density along that sightline) and a Gaussian width of $\sigma = 2$, resulting in $\chi^{2}=0.76$. Considering our method is an iterative procedure it produces models that approach the data successively closer with each iteration. On the other hand, the ideal value for $\chi^{2}$ is unity, which would indicate that the measured densities and model are in agreement within the error variance. However, in cases where the errors on the data have been overestimated the $\chi^{2}$ would be less than unity. Actually, to prevent the algorithm from getting stuck in local minima instead of the global minimum, we deliberately increased the noise on the measured column densities. Therefore, our $\chi^{2}$ is a little below unity. If the $\chi^{2}$ value did not change in a subsequent iteration the algorithm was considered to have converged. In previous studies, the reduced $\chi^{2}$ for Na\,{\sc i} was reported to be $1.5$, and $1.8$ for H\,{\sc i}\cite{2010A&A...510A..54W}. We varied the Gaussian width in the range of $\sigma \pm 2$ and found exactly the same results.

\subsection{Approach}
There are two approaches to solve equation\,\ref{eq:e1}. The first method is to divide the entire 3D space into equal voxels and estimate the model parameters in every single voxel. Another way is to estimate the model parameters along each sightline without dividing the space into separate voxels. In the first method, the un-smoothed distribution of stars within space, and the low number of observed sightlines in the case of small voxels cause some voxels not to be crossed by any star sightline. Therefore the number of constraints would be less than the number of voxels. In that case, the iteration will not converge to a correct answer. This problem can be circumvented by increasing the size of the voxels so as to put at least one star within every single voxel, to maximize the number of constraints. This is done at the expense of map resolution, which can result in the oversight of fine structures. The second method is the non-blocking method\cite{1982RvGSP..20..219T}, which, instead of dividing the entire volume into voxels, only divides each observed sightline into segments and then estimates the volume density at every single point by inversion. In our case, we divided each sightline into $30$ segments while for the furthest point in a $540$ pc box we could estimate the volume density at each $30$ pc. Although most target stars are scattered at distances below $150$ pc, allowing a volume density estimate at each $5$ pc, by dividing the 3D space into $30^{3}$ pc voxels the final volume density in each voxel is the weighted mean value of all densities passing through the voxel. The weighted mean density in each voxel is computed based on the following equation:
\begin{equation}
    N_{v}=\frac{ \sum _{i=1}^{m}\frac{l_{i}}{D}N_{i}}{ \sum _{i=1}^{m}\frac{l_{i}}{D}}
\end{equation}
Here, $N_{i}$ and $l_{i}$ are the volume density and length of i-th segment, respectively, and $D$ is the space diagonal of the voxel. In cases where a zero column density sightline passes through a voxel crossed by some other non-zero column densities, the volume density within the voxel decreases. In the returned DIB map there are some strong absorptions seen through regions with low volume density; in simplistic map-making methods, these sightlines would cause the ``fingers of God'' radial striping.

Considering the correlation function represents the relation of each pair of points, the convergence of the algorithm sensitively depends on the acceptable fluctuation level of the model around the best estimate ($\sigma_{m}$). This value depends on the distribution and quantity of the data and is determined based on trial and error. To obtain the optimal value for $\sigma_{m}$ we tested different values and found that by using $\sigma_{m}=0.08$ the inversion converges to the minimum $\chi^{2}=0.76$ in only 5 iterations. If we choose a larger value for $\sigma_{m}$ the algorithm may diverge in the first iterations, while by picking very small values for $\sigma_{m}$ the process may need more than $100$ iterations before converging which in the case of non-linear inversions is prohibitive. After converging to an acceptable $\chi^{2}$ value, if the estimated column density was less than $0.5 N_{obs}$ or greater than $2.5 N_{obs}$ we considered it an outlier\cite{2010A&A...510A..54W}. Additionally, we chose a very low a priori density, $\rho_{0} = 10^{-10}$ cm$^{-3}$, which is one dex below the a priori value for corresponding Na\,{\sc i} atoms.

To check the reliability of the inversion, we simulated the result obtained by applying the method on an artificial distribution of clouds which is described in the Supplementary file.

\subsection{Test the inverse method}
To check the reliability of the inversion, we simulated the result obtained by applying the method on an artificial distribution of clouds. In this simulation we produced three rectangular clouds with different densities and separated from one another in order to check the inversion’s ability to determine the clouds’ edges. In the first test, we placed multiple stars in each voxel. In this case the number of constraints exceeds the number of voxels and the returned map is similar to original clouds in term of distribution and densities. We then estimated the cloud densities by using randomly distributed stars, fewer in number than voxels. The lack of constraints as a result of empty voxels renders the inversion method unable to accurately determine the shape of the clouds – especially at the cloud edges (See Supplementary Fig.\,6). As another test, we randomly varied the observed column densities within their error range (based on a random Gaussian distribution) and applied the inversion method on the new dataset. The returned results were, in essence, identical to those from the previous test, and the map did not show significant differences. The outer layers of clouds showed very small changes which were, however, smoothed out once the final voxel median smoothing had been applied.

\end{methods}

\renewcommand{\figurename}{Supplementary Figure}
\setcounter{figure}{0}

\clearpage

\begin{figure*}
\centering
\includegraphics[width=14cm]{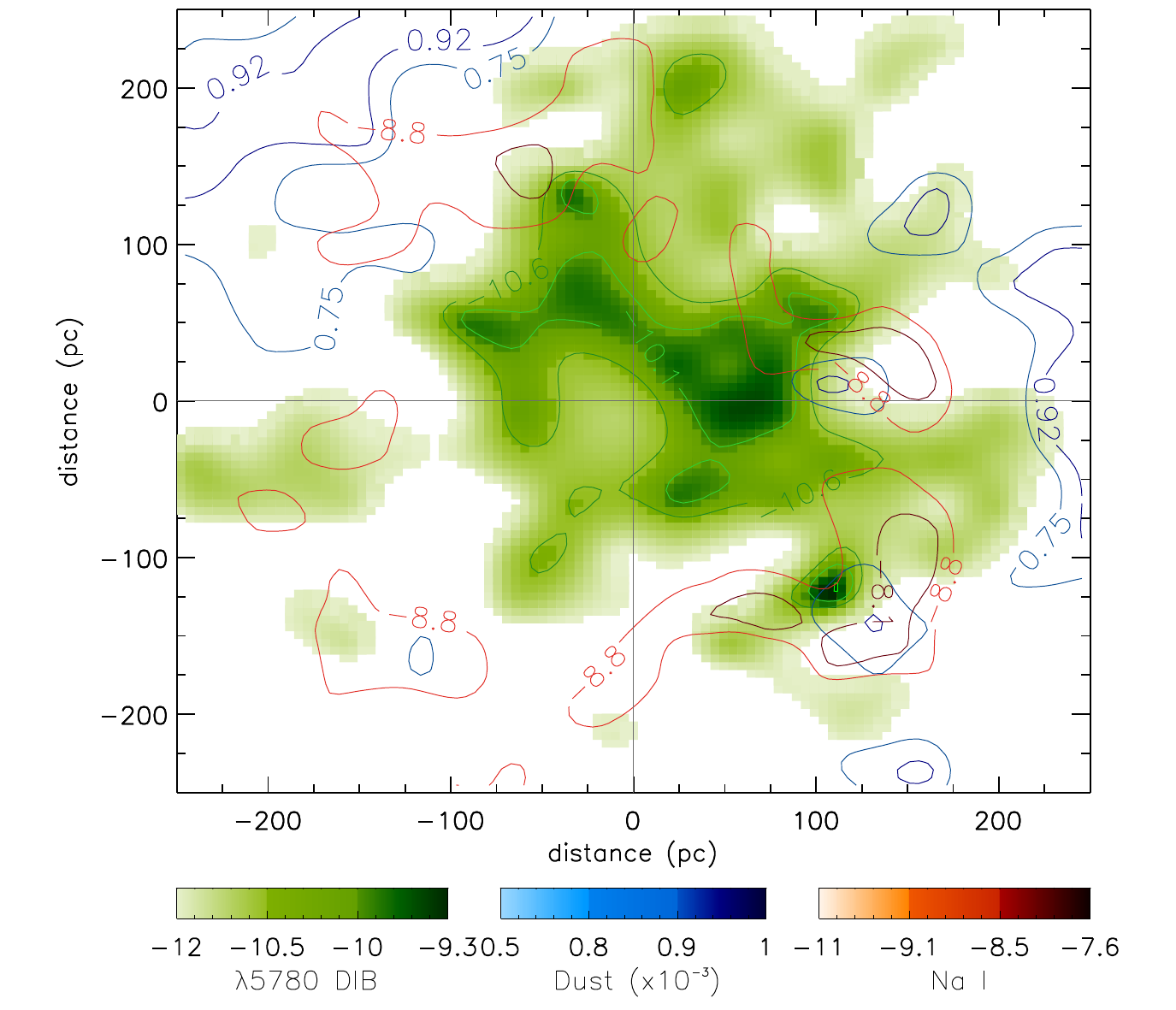}
\caption{\textbf{GP map in three different ISM tracers}. The red (R) channel represents the Na\,{\sc i} distribution\cite{2010A&A...510A..54W}, whilst the green (G) channel is the $\lambda5780$ DIB distribution; the color scales are logarithmic volume densities. The blue (B) channel represents the dust distribution\cite{2014A&A...561A..91L}; its color scale represents the $E(B-V)$/pc distance-normalized photometric reddening.
}
\end{figure*}
\clearpage

\begin{figure*}
\centering
\includegraphics[width=14cm]{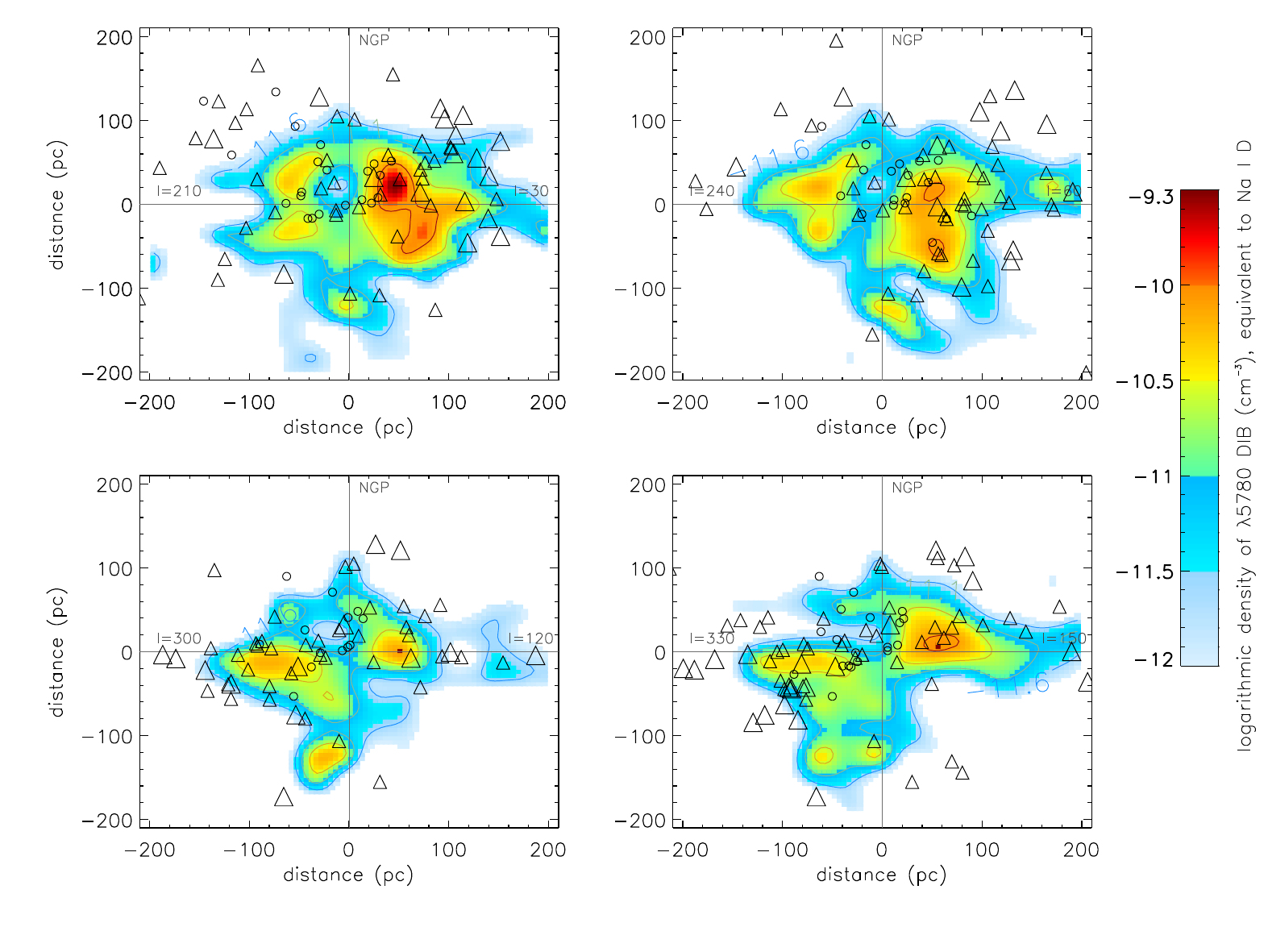}
\caption{\textbf{$\lambda5780$ DIB vertical slices}. Plane $l=30\degree$ (Upper-left panel): the map shows a dense DIB concentration in the direction of $l=30\degree$ (i.e. to the right) at the position of a compact dust cloud\cite{2014A&A...561A..91L} stretched between $50$ pc underneath the GP up to $50$ pc above it. Plane $l=60\degree$ (Upper-right panel): the density is still highest in the direction of $l=60\degree$ around the GP; the structure in the opposite direction of $l=240\degree$ at a distance of $80$ pc is clearly separated from this by a vacuous tunnel. Plane $l=120\degree$ (lower-left panel): most of the DIB carrier material is found in the direction of $l=300\degree$ (i.e. to the left), in front of -- but not within -- the Coalsack, Chamaeleon and Musca clouds residing at distances of $120$--$250$ pc in and below the GP\cite{2014A&A...561A..91L}. Plane $l=150\degree$ (lower-right panel): a DIB cloud is located toward $l=150\degree$ between $50$--$100$ pc in the direction of a very dense dust cloud\cite{2014A&A...561A..91L}. In the opposite direction of $l=330\degree$ widespread DIB material is detected in front of the Lupus H\,{\sc i} cloud, Lupus supernova remnant and dust clouds.
}
\end{figure*}
\clearpage

\begin{figure*}
\centering
\includegraphics[width=14cm]{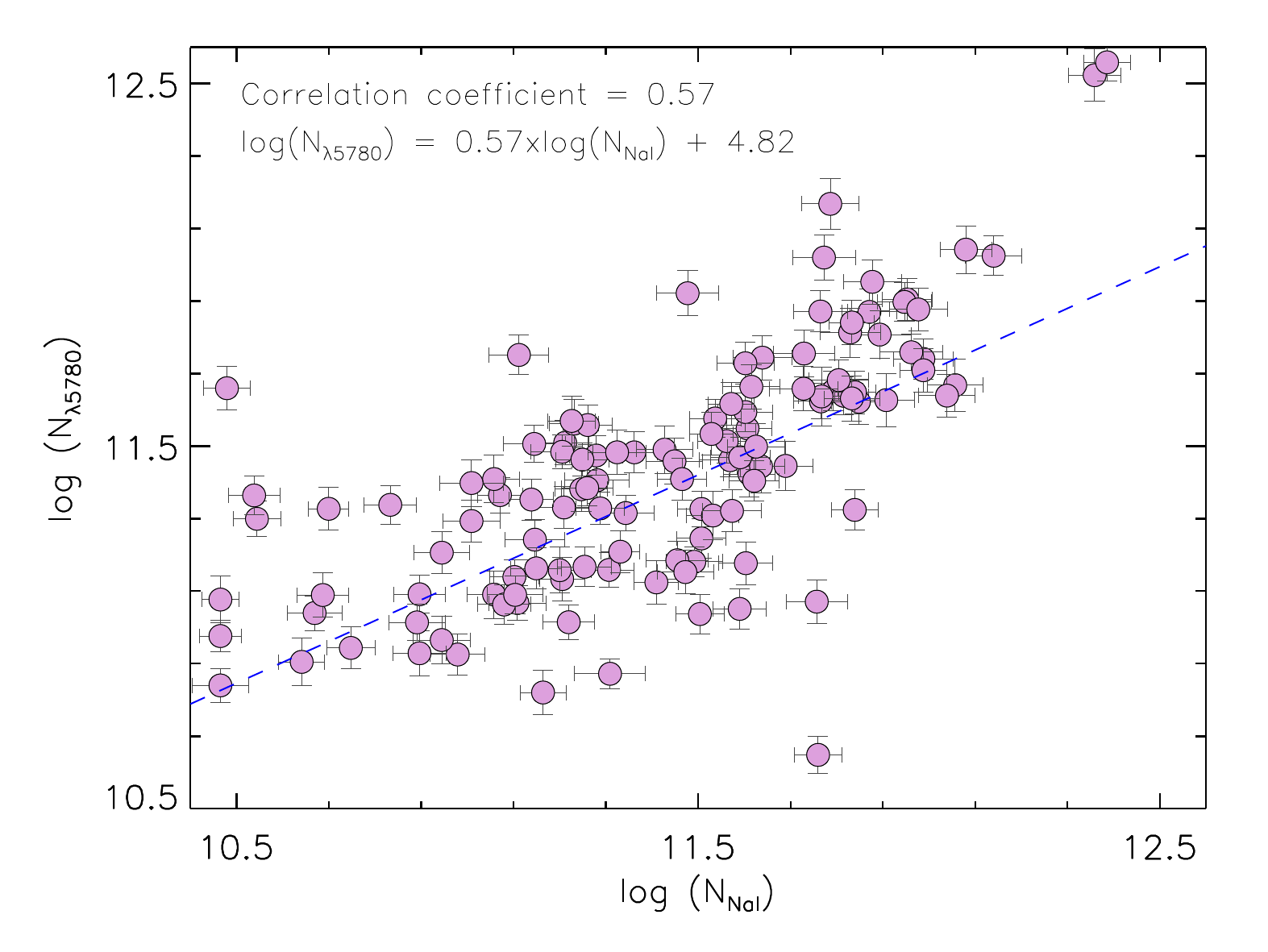}
\caption{\textbf{Logarithmic $\lambda5780$ DIB column densities versus logarithmic Na\,{\sc i} column densities}. In order to be able to assign values to the densities we derive, the values for the oscillator strengths of the DIBs have been fixed to bring the densities in line with those determined for Na\,{\sc i} (which is a good tracer of the overall gas density). The best correspondence is achieved for a value of $f=0.3$, hence forthwith we adopt. The graph shows that the relation breaks down at low values, where the $\lambda5780$ DIB seems to reach a plateau while the Na\,{\sc i} still decreases. This could be associated with the rarefied hot ISM component, which is better traced by the $\lambda5780$ DIB than by Na\,{\sc i} and the $\lambda5797$ DIB. While this modifies our results in a quantitative sense, it does not change the morphology of our maps and only corroborates our findings with regard to the resilience of the $\lambda5780$ DIB carrier within the LB. The dashed blue line is a Theil--Sen regression fit with a correlation coefficient of $0.57$. Error bars determine the logarithm of 1-$\sigma$ uncertainty of the column densities.
}
\end{figure*}
\clearpage

\begin{figure*}
\centering
\includegraphics[width=14cm]{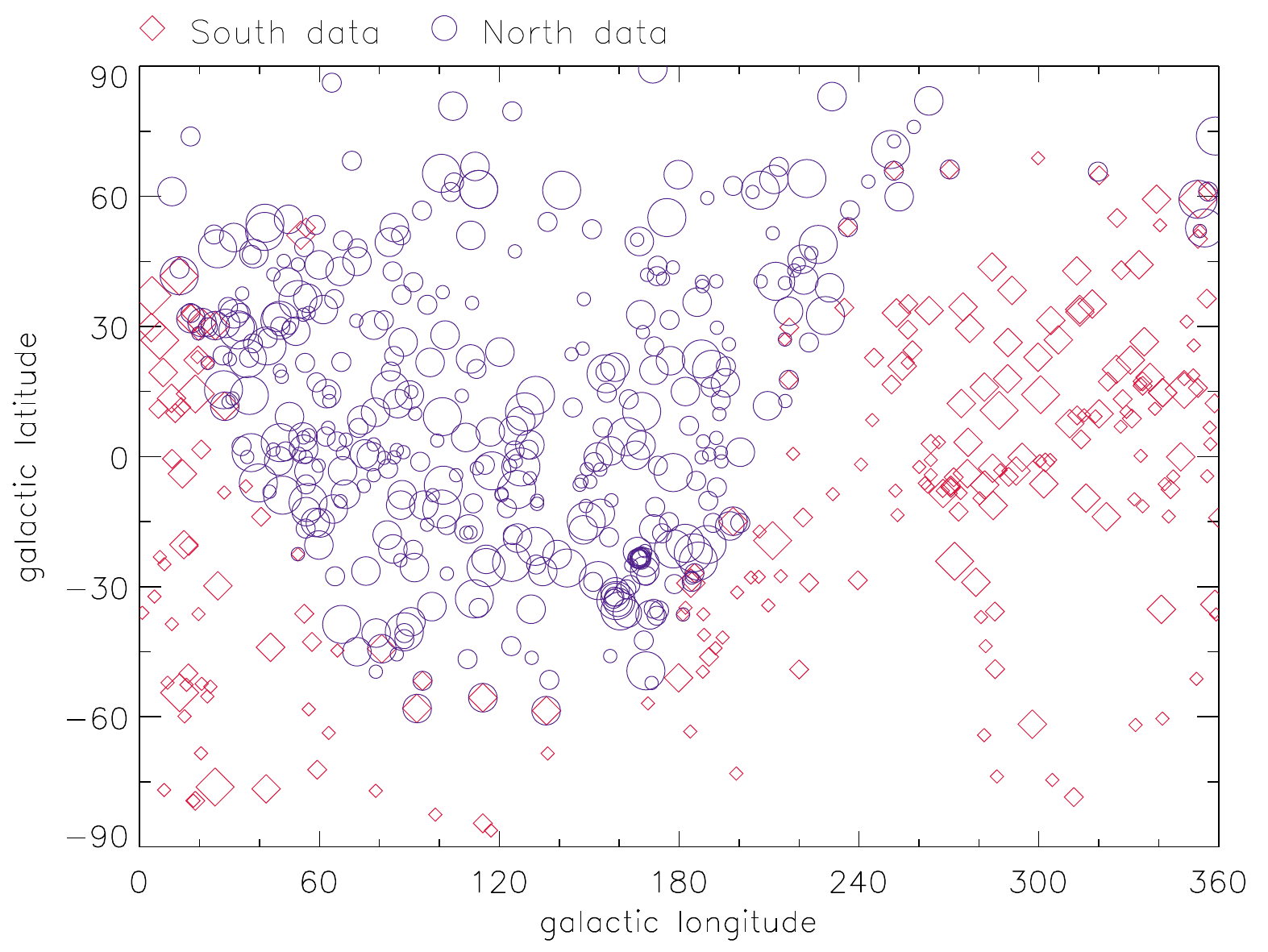}
\caption{\textbf{Galactic latitude versus Galactic longitude of the observed sightlines}. The blue circles are the Northern target stars and the red circles the Southern ones. The biggest circles correspond to distances less than $50$ pc from the Sun while the smallest correspond to distances beyond $150$ pc.
}
\end{figure*}
\clearpage

\begin{figure*}
\centering
\includegraphics[width=14cm]{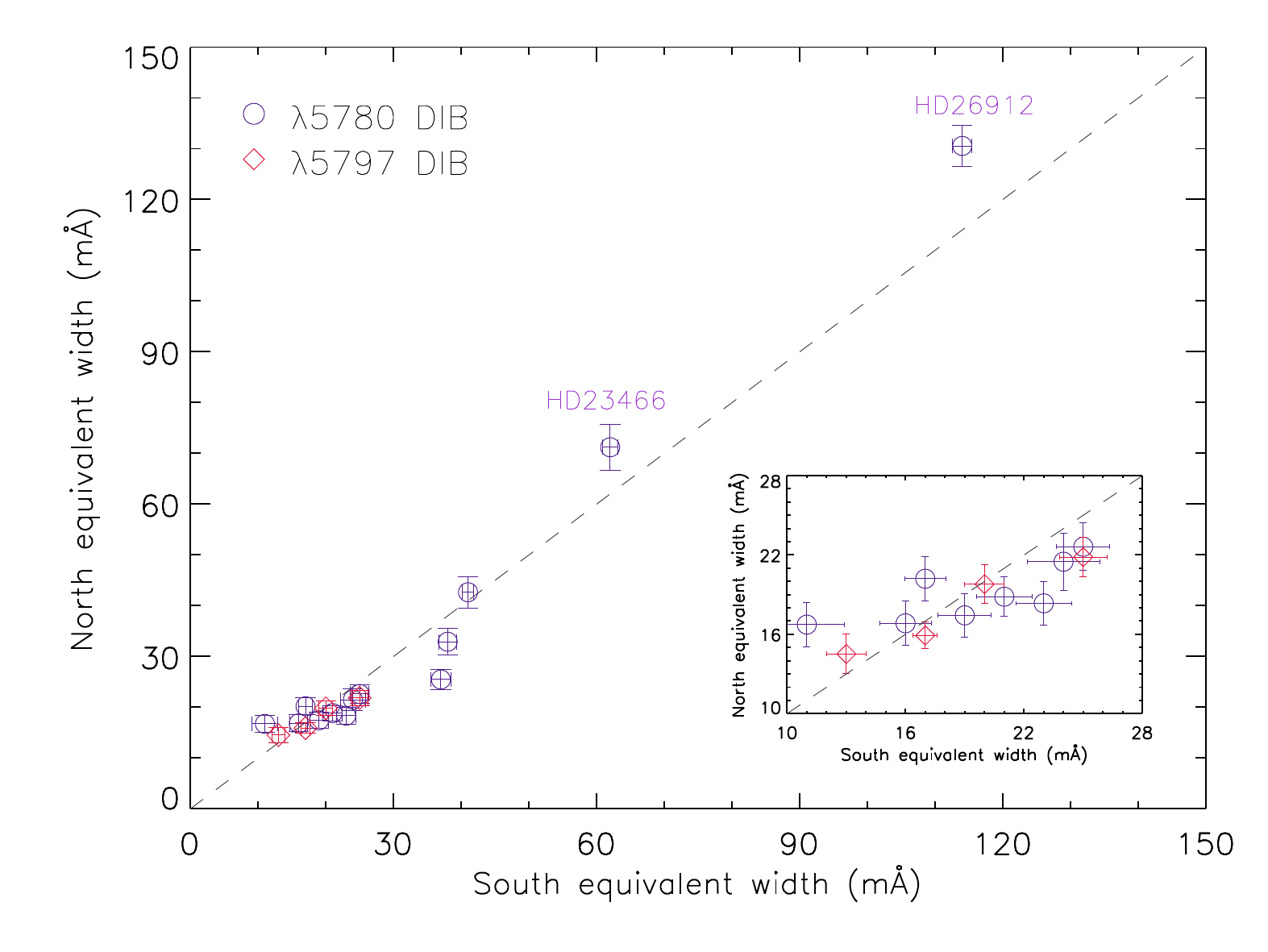}
\caption{\textbf{INT measured equivalent width versus NTT measured equivalent width}. Amongst the Northern and Southern samples, $15$ stars were in common. To check the quality of our measurements with the two different telescopes and instruments, we compared the two values for the equivalent width of the same DIB. The blue circles represent the $\lambda5780$ DIB measurements and the red circles the $\lambda5797$ DIB measurements. The correspondence is overall satisfactory. The zoomed lower left corner of plot is overplotted in figure. In this comparision, the relative uncertainties on the EWs are 2--19\% of the measured values in both observations. Also, the measured values are deviated from the dashed line by the root mean square of 3.28. However, in general, the mean value of EWs in our survey is EW$\sim$30$\pm$15\%. Error bars determine 1-$\sigma$ uncertainty of the equivalent widths.
}
\end{figure*}
\clearpage

\begin{figure*}
\centering
\includegraphics[width=16cm]{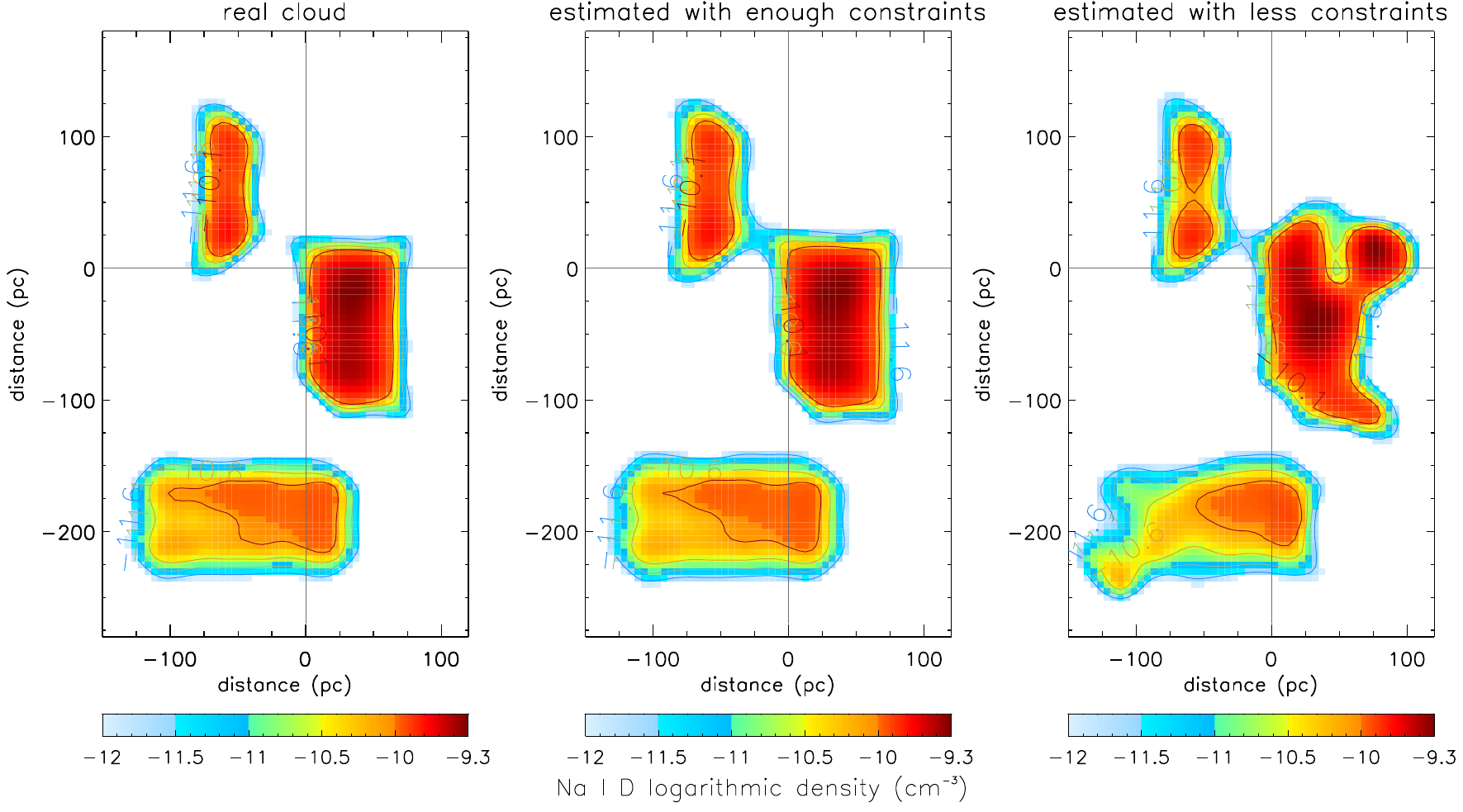}
\caption{\textbf{Simulated and re-constructed map with our inverse method}. The left panel is a simulated cloud in the GP view. This map comprises three clouds, colored based on their $\log n_{\lambda}$, and with three levels for logarithmic volume densities of $-9.6$, $-10.5$, and $-11.5$ outlined with contours. The middle panel shows the map as determined with enough constraints. In this case every voxel contains at least one star, therefore the number of constraints exceeds the number of voxels. The returned map in this case is undistinguishable from the truth. The right panel shows the map as determined with fewer constraints than voxels. Therefore, the inverse method could not recover the true cloud shapes and densities. It shows that for an accurate map, the number of observed sightlines should be as large as the number of voxels.
}
\end{figure*}
\clearpage

\clearpage
\begin{addendum}

\item[Data availability]
The data that support the plots within this paper and other findings of this study are available from the corresponding author upon reasonable request.

\end{addendum}

\end{document}